\documentclass[final,5p,times,twocolumn]{elsarticle}

\usepackage{amsmath,amssymb,lineno}
\usepackage[hyphens]{url}
\usepackage[colorlinks]{hyperref}
\AtBeginDocument{%
  \hypersetup{%
    linkcolor=red,%
    citecolor=blue,%
  }%
}%
\usepackage{graphicx}

\biboptions{sort&compress}

\newcommand\araa{ARA\&A~}
\newcommand\apj{ApJ~}
\newcommand\apjl{{ApJL~}}
\newcommand\apjs{{ApJS~}}
\newcommand\apss{{Ap\&SS~}}
\newcommand\aap{{A\&A~}}
\newcommand\mnras{{MNRAS~}}
\newcommand\prd{{Phys. Rev.~D}}
\newcommand\pasj{{PASJ~}}
\newcommand\nat{{Nature~}}
\newcommand\nar{{New~Astronomy~Review~}}

\journal{Astroparticle Physics}

\begin{document}

\begin{frontmatter}
\title{High Energy Gamma Rays from Nebulae Associated with Extragalactic Microquasars and Ultra-Luminous X-ray Sources}

\author[isas]{Yoshiyuki Inoue}
\ead{yinoue@astro.isas.jaxa.jp}

\author[isas,kyoto]{Shiu-Hang Lee}
\ead{herman@kusastro.kyoto-u.ac.jp}

\author[hiroshima]{Yasuyuki T. Tanaka}
\ead{ytanaka@hep01.hepl.hiroshima-u.ac.jp}

\author[tokyo]{Shogo B. Kobayashi}
\ead{kobayashi@juno.phys.s.u-tokyo.ac.jp}

\address[isas]{Institute of Space and Astronautical Science JAXA, 3-1-1 Yoshinodai, Chuo-ku, Sagamihara, Kanagawa 252-5210, Japan}
\address[kyoto]{$^2$Department of Astronomy, Kyoto University, Kitashirakawa, Sakyo-ku, Kyoto 606-8502, Japan}
\address[hiroshima]{$^3$Hiroshima Astrophysical Science Center, Hiroshima University, 1-3-1 Kagamiyama, Higashi-Hiroshima, Hiroshima 739-8526, Japan}
\address[tokyo]{Department of Physics, The University of Tokyo, 7-3-1 Hongo, Bunkyo-ku, Tokyo 113-0033, Japan}

\begin{abstract}
In the extragalactic sky, microquasars and ultra-luminous X-ray sources (ULXs) are known as energetic compact objects locating at off-nucleus positions in galaxies. Some of these objects are associated with expanding bubbles with a velocity of 80--250 ${\rm km~s^{-1}}$. We investigate the shock acceleration of particles in those expanding nebulae. The nebulae having fast expansion velocity $\gtrsim120~{\rm km~s^{-1}}$ are able to accelerate cosmic rays up to $\sim100$~TeV. If 10\% of the shock kinetic energy goes into particle acceleration, powerful nebulae such as the microquasar S26 in NGC~7793 would emit gamma rays up to several tens TeV with a photon index of $\sim2$. These nebulae will be good targets for future Cherenkov Telescope Array observations given its sensitivity and angular resolution. They would also contribute to $\sim7$\% of the unresolved cosmic gamma-ray background radiation at $\ge0.1~{\rm GeV}$. In contrast, particle acceleration in slowly expanding nebulae $\lesssim120~{\rm km~s^{-1}}$ would be less efficient due to ion-neutral collisions and result in softer spectra at $\gtrsim10$~GeV.
\end{abstract}


\end{frontmatter}


\section{Introduction}
\label{sec:intro}
In the very high energy (VHE; $\gtrsim50$~GeV) gamma-ray sky, a few hundreds of objects have been detected by the Large Area Telescope (LAT) on board the {\it Fermi} gamma-ray telescope ({\it Fermi}) \cite{ack16_2fhl} and by the imaging atmospheric Cherenkov telescopes \cite[see e.g.][]{wak08}. Further progress is anticipated in the near future by the Cherenkov Telescope Array (CTA) \cite[][]{act11}. Its improved flux sensitivity and angular resolution will enable us to unveil new particle accelerators in the Universe.

In the extragalactic sky, various source classes have been considered for future CTA observations such as active galactic nuclei \citep{sol13}, gamma-ray bursts \citep{ino13_grb}, star forming galaxies \citep{ace13}, and cluster of galaxies \citep{ace13}. Here, the angular resolution of CTA will achieve $\lesssim3$~arcmin at $\gtrsim1$~TeV\footnote{\url{https://portal.cta-observatory.org/Pages/CTA-Performance.aspx}}. As angular sizes of nearby galaxies up to $\sim10$~Mpc is $\sim10$~arcmin, it would be possible to spatially resolve particle accelerators in those galaxies. 

It would be difficult to detect supernova remnants or pulsar wind nebulae in extragalactic galaxies considering their luminosities ($L_\gamma\lesssim10^{36}~{\rm erg~s^{-1}}$). Here, some galaxies are known to host more powerful compact objects such as microquasars and ultra-luminous X-ray sources (ULX) whose kinetic or radiative power is $\gtrsim10^{39}~{\rm erg~s^{-1}}$. These objects are discovered at off-nucleus positions \citep{fen11}. 

Microquasars are X-ray binary systems having relativistic bipolar outflows or jets whose kinetic power is greater than $10^{39}$~erg s$^{-1}$ \citep{mir99}. ULXs are also compact X-ray binary systems having X-ray luminosities greater than $10^{39}$~erg s$^{-1}$ \citep{fen11}. Although X-ray emission mechanisms from microquasars and ULXs are not fully understood yet \footnote{There are three distinct ideas widely considered to interpret high X-ray luminosities of ULXs, although there is no general agreement on their nature; sub- or trans-Eddington accretion onto intermediate mass black holes with a mass of $M_{\rm BH}\gg 10 M_\odot$ \citep[e.g.][]{mak00}, supercritical mass accretion onto stellar mass black holes \citep[e.g.][]{gla09,kaw12}, or accreting pulsars \citep[e.g.][]{bac14}.  }, some of them are known to be associated with expanding nebulae with a velocity of $80-250~{\rm km\ s^{-1}}$ and a size of $\sim200~{\rm pc}$ \citep{pak03,cse12}. Since the kinetic power of those nebulae is known to be comparable to or even greater than the radiative or jet kinetic power \citep{cse12}, the nebulae are good candidates as new cosmic-ray acceleration sites and possibly they would emit gamma rays through hadronuclear interactions with ambient gases. 

In this paper, we investigate the diffusive shock acceleration of particles in the expanding nebulae associated with extragalactic microquasars and ULXs. And, we estimate expected gamma-ray and neutrino signals from those nebulae. We further consider their contribution to the cosmic gamma-ray and neutrino background radiation and future detectability by CTA. Throughout this paper, we define $Q_x=Q/10^x$.

\section{X-ray Source Embedded Bubbles in the local Universe}
The size of the microquasar and ULX bubbles $R_b$  is of an order of 200~pc and the expansion velocity of the bubbles is known to be $v_s=80-250~{\rm km\ s^{-1}}$ \citep{pak03,cse12}. Following the self-similar expansion law \citep{wea77,kai97}, the bubble size can be described as
\begin{equation}
R_b\approx0.76(P_{\rm kin}/\mu m_p n_{\rm gas})^{1/5}t^{3/5}\sim200 P_{{\rm kin},40.5}^{0.2}n_{{\rm gas}, 0.5}^{-0.2}t_{13.5}^{0.6}~{\rm pc},
\end{equation}
where $P_{\rm kin}$ is the time-averaged kinetic power, $\mu=0.61$ is the mean molecular weight, $m_p$ is the proton mass, $n_{\rm gas}$ is the gas density, and $t$ is the age of the system. Thus, the characteristic age of the bubble is 
\begin{equation}
\tau = 3R_b/5v_s \sim 4.9\times10^5 R_{b, 20.8}v_{s, 7.1}^{-1}~{\rm yr}.
\end{equation}
This age is comparable to the expected ULX lifetime \citep[e.g.][]{min12,ino16}. Once we measure the bubble size and the expanding velocity, the time-averaged kinetic power of the bubble can be described as
\begin{equation}
P_{\rm kin}\approx 18 \mu m_p n_{\rm gas} R_b^2 v_s^3 \sim 3.6\times10^{40}R_{b, 20.8}^2v_{s, 7.1}^3 n_{{\rm gas}, 0.5}~{\rm erg~s^{-1}}.
\end{equation}

The Mach number of the shock due to the expanding nebulae is estimated as 
\begin{equation}
\label{eq:Mach}
\mathcal{M}\approx v_s/(\gamma k_B T_u/\mu m_p)^{1/2}\sim 8.0v_{s,7.1}T_{u,4}^{-1/2},
\end{equation}
where $\gamma=5/3$ is the specific heat ratio, $k_B$ is the Boltzmann constant, and $T_u$ is the temperature of the upstream of the shock. $\mathcal{M}$ is larger than unity for the upstream cold gas component with $T_u<10^4$~K. The dynamical timescale in which the shock dissipates is
\begin{equation}
\label{eq:t_dyn}
t_{\rm dyn}\approx \frac{R_{\rm b}}{v_s}\sim 8.1\times10^5 R_{b, 20.8}v_{s, 7.1}^{-1}~{\rm yr}.
\end{equation}

The downstream temperature is estimated as 
\begin{equation}
\label{eq:T_d}
T_d\approx2(\gamma-1)\mu m_p v_s^2/k_B/(\gamma+1)^2\sim2.0\times10^5 v_{s,7.1}^2 ~{\rm K}.
\end{equation}
Thus, radiative cooling plays an important role. For a solar metallicity interstellar medium (ISM) environment, the cooling time scale is given by \citep[Eq. 34.4 of][]{dra11}
\begin{equation}
\label{eq:t_rad}
t_{\rm rad}\sim 1.4\times10^3 n_{\rm gas,0.5}^{-1}v_{s,7.1}^{17/5}~{\rm yr},
\end{equation}
for $10^5\lesssim T\lesssim10^{7.3}$~K. If the metallicity is $0.1Z_\odot$, the time scale will become a factor of 3 longer. From \autoref{eq:t_dyn} and \autoref{eq:t_rad}, the downstream plasma cools within dynamical timescale. In other words, these nebula shocks are radiative. UV photons from the radiative zone would significantly ionize the upstream ISM. At $v_s\gtrsim120~{\rm km~s^{-1}}$, 
shock induced ionizing radiation is strong enough to completely ionize the upstream gas \citep{shu79,hol89,lee15}. At the downstream, the temperature decreases and the density increases as the radiation cools the gas in the downstream. As the shock compression ratio $r = (\gamma+1)\mathcal{M}^2 / [(\gamma-1)\mathcal{M}^2 + 2]\sim4$ in our case, we assume the spectral index of 2 for simplicity in this paper.

Here, the magnetic field in the shock upstream can be described as
\begin{equation}
B_u = 1.7b\sqrt{n_{{\rm gas}, 0.5}}~{\rm \mu G},
\end{equation}
where the dimensionless parameter $b$ is $\sim 1$ from Zeeman measurements of self-gravitating molecular clouds in the Galaxy \citep{cru99}. The magnetic field in the downstream region will be amplified by compression as
\begin{equation}
\label{eq:B_d}
B_d = \sqrt{\frac{1}{3}+\frac{2}{3}r^2} B_u \sim 5.7 n_{{\rm gas}, 0.5}^{1/2}~{\rm \mu G}.
\end{equation}

Here, if Alfv\'en-wave turbulence is fully generated, $B_d$ can be amplified locally around the shocks \citep[see e.g.][]{ski75,bel78a,bel78b,bel04}. In such cases, $B_d$ would become as high as $B_d\approx(4\pi\xi_Bf_{\rm ion}\mu n_{\rm gas}m_pv_s^2)^{1/2}\sim74\xi_{B,0}^{1/2}f_{{\rm ion}, 0}^{1/2}n_{{\rm gas}, 0.5}^{1/2}v_{s,7.1}~{\rm {\mu}G}$, where $\xi_B$ is the magnetic field amplification factor and $f_{\rm ion}\equiv n_n/(n_n+n_i)$ is the ionization fraction. $n_n$ and $n_i$ is the neutral particle density and the ionized particle density, respectively. However, we note that ion-neutral collisions suppress Alfv\'en wave \citep{shu79,lee15} at $v_s\lesssim120~{\rm km~s^{-1}}$. In this paper, we take \autoref{eq:B_d} as the fiducial value for the downstream magnetic field. 

The diffusive shock acceleration timescale can be written as
\begin{equation}
t_{\rm acc}\approx\frac{10}{3}\frac{\eta c r_g}{v_s^2}\sim1.3\times10^6Z^{-1}n_{{\rm gas}, 0.5}^{-1/2}v_{s,7.1}^{-2}E_{{\rm cr}, 14}~{\rm yr},
\label{eq:t_acc}
\end{equation}
where $r_g=E_{\rm cr}/ZeB_d$ is the gyroradius and $\eta\ge1$. For simplicity, we take $\eta=1$ in this paper, i.e. the Bohm limit, since ULXs are known to be associated with star forming regions \citep[e.g.][]{swa09,gri11,pou13} which would amplify the upstream turbulence. Such an efficient acceleration is possibly seen in a Galactic supernova remnant by assuming a simple diffusive shock acceleration model \citep{uch07}, although $\eta$ can be different from unity by considering escape-limited acceleration \citep{ohi10,hes16} or stochastic acceleration \citep{fan10}.

The attainable maximum cosmic-ray energy is given by $t_{\rm acc}=t_{\rm dyn}$ (See \autoref{eq:t_dyn} and \autoref{eq:t_acc}):
\begin{eqnarray}
E_{\rm cr, max}&=&\frac{3Z eB_dR_{\rm b}v_s}{10\eta c}\\
&\sim&1.3\times10^{14} \eta_{0}n_{{\rm gas}, 0.5}^{1/2}v_{s,7.1}R_{{\rm b}, 20.8}~{\rm eV}, \label{eq:Emax}
\end{eqnarray}
TeV cosmic rays can be affordable from the extragalactic X-ray source nebulae. If the Alfv\'en-wave turbulence is fully generated, it will reach to $\sim1$~PeV.

The ionization state in the shock downstream strongly depends on its velocity \citep{shu79,hol89,lee15}. If neutral particle exists, i.e. $v_s\lesssim120~{\rm km~s^{-1}}$, ion-neutral collisions will lead damping of the magnetic turbulence in the shock precursor and those collisions will hamper acceleration of particles at the highest energies. The expected break energy can be estimated as \citep[][and references therein]{mal11}
\begin{equation}
E_{\rm cr, br}\sim 1.9\times10^{9}~T_{u,4}^{-0.4}B_{u,-6}^2(1-f_{\rm ion})^{-1}f_{\rm ion}^{-1/2}n_{{\rm gas}, 0.5}^{-3/2}~{\rm eV}.
\end{equation}
Thus, at $v_s\lesssim120~{\rm km~s^{-1}}$, the particle spectrum will have a break at $\sim10$~GeV and become steeper by one power above the break. The ion fraction $f_{\rm ion}$ is estimated according to the steady-state model by \citet{hol89}.

Let us consider energy loss processes for cosmic rays. The energy loss timescale due to $pp$ interactions can be estimated as
\begin{equation}
t_{pp}\approx\frac{1}{\kappa_{pp}4n_{\rm gas}\sigma_{pp}c}\sim 6.0\times10^{6} n_{{\rm gas},0.5}^{-1}~{\rm yr},
\end{equation}
where the cross section $\sigma_{pp}\sim3\times10^{-26}~{\rm cm^2}$  for $10^{10-12}$~eV \citep{kam06,kaf14} and  the inelasticity $\kappa_{pp}\approx0.5$. The factor of 4 in front of $n_{\rm gas}$ is due to compression in the downstream. We note that the $pp$ cross section depends logarithmically on the proton energy.

By considering the steady-state for simplicity, the hadronuclear interaction efficiency is estimated as
\begin{equation}
f_{\rm pp}=\frac{t_{\rm dyn}}{t_{\rm pp}}\sim0.28 R_{{\rm b}, 20.8}v_{s, 7.1}^{-1}n_{{\rm gas},0.5}.
\end{equation}
In hadronuclear interactions, charged and neutral pions are generated at the ratio of $\pi^+:\pi^0\approx2:1$. Gamma rays are produced by the decay of neutral pions as $\pi^0\rightarrow2\gamma$.

If we neglect ion-neutral collisions, gamma-ray flux from an expanding nebula can be estimated as
\begin{eqnarray}
E_\gamma^2\frac{dN_\gamma}{dE_\gamma}&\approx& \frac{1}{3} \frac{\epsilon_{\rm M}\xi_{\rm cr}P_{\rm kin}{\rm min}[1, f_{pp}]}{4\pi d^2\mathcal{C}},\\ \nonumber
&\sim&1.1\times10^{-13} ~{\rm erg~cm^{-2}~s^{-1}}~\left(\frac{\mathcal{C}}{12}\right)^{-1}\left(\frac{d}{2~{\rm Mpc}}\right)^{-2}\\
&\times&\xi_{{\rm cr},-1}R_{{\rm b}, 20.8}^3v_{s, 7.1}^{2}n_{{\rm gas},0.5}^2
\end{eqnarray}
where $\xi_{\rm cr}$ is the cosmic-ray acceleration efficiency and $\mathcal{C}=\ln(E_{p,{\rm max}}/E_{p, {\rm min}})$. We set $E_{p, {\rm max}}=10^{14}$~eV and $E_{p,{\rm min}}=10^9$~eV. We set the nuclear enhancement factor $\epsilon_M=1.85$ as in our Galaxy \citep[e.g.,][]{ste70,mor09}. ULXs are known to be hosted by low-metallicity galaxies at the metallicity of $Z\lesssim 1/2Z_\odot$ \citep{map10}. Thus, $\epsilon_M$ would be lower than that in our galaxy. However, it is known that a galaxy is not chemically homogeneous and can have metallicity dispersion by a factor of 10 in a galaxy \citep[e.g.][]{rol00,nii11}. We note that an X-ray measurement of the ULX NGC~1313 X-1 revealed that the local oxygen abundance is $\sim$50\% of the solar value using the low energy X-ray absorption feature \citep{miz07}.

\begin{table}[t]
\begin{center}
\caption{Observed Properties of Nebulae Associated with Extragalactic Compact X-ray Objects \label{tab:neb}}
	\begin{tabular}{lcccc}
	\hline
	{Object} & {$d$} & {$L_X$}  & {$R_b$} & {$v_{\rm s}$}\\ \hline
	{} & {(Mpc)}  & {$({\rm erg~s^{-1}})$} & {$({\rm pc})$} & {$({\rm km~s^{-1}})$}\\ \hline
	Microquasar S26 & 3.9 & $6.2\times10^{36}$ & 200 & 250 \\
	ULX IC~342~X-1 & $3.9$ & $1.6\times10^{40}$  & 200 & 100\\
	HMXB IC~10~X-1 & 0.79 & $1.5\times10^{38}$   & 170 & 80\\
	\hline
	\end{tabular}
\caption{The table data are taken from \citet[][for S26]{pak10}, \citet[][for IC~342~X-1]{all08,cse12}, and \citet[][for IC~10~X-1]{loz07}. The data for the time-averaged X-ray luminosity are from \citet{cse12}.}
\end{center}
\end{table}

We can also evaluate the expected neutrino flux per flavor as $E_\gamma dN_\gamma/dE_\gamma \approx 2 \times E_\nu dN_\nu/dE_\nu (E_\nu=0.5 E_\gamma)$, or
\begin{eqnarray}
 \nonumber
E_\nu^2\frac{dN_\nu}{dE_\nu}&\sim&5.6\times10^{-14} ~{\rm erg~cm^{-2}~s^{-1}}~\left(\frac{\mathcal{C}}{12}\right)^{-1}\left(\frac{d}{2~{\rm Mpc}}\right)^{-2}\\
&\times&\xi_{{\rm cr},-1}R_{{\rm b}, 20.8}^3v_{s, 7.1}^{2}n_{{\rm gas},0.5}^2.
\end{eqnarray}

\begin{figure}
 \begin{center}
  \includegraphics[width=8.5cm]{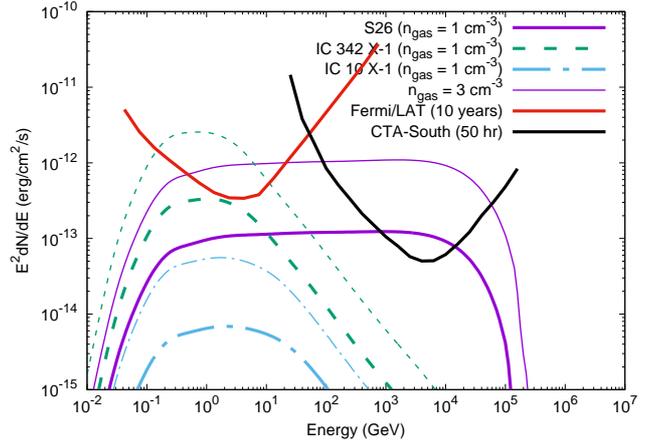} 
 \end{center}
\caption{Expected gamma-ray spectra from nebulae associated with microquasar S26, ULX IC~342~X-I, and HMXB IC~10~X-I shown by solid, dashed, and dot-dashed curve, respectively. Thick curves represent the case with $n_{\rm gas}=1~{\rm cm^{-3}}$, while thin curves show $n_{\rm gas}=3~{\rm cm^{-3}}$. The sensitivity of {\it Fermi}/LAT and CTA in the southern site is shown for 10~yr and 50~hr integration, respectively.}\label{fig:nebula}
\end{figure}

\autoref{fig:nebula} shows the expected gamma-ray signals from nebulae hosting compact X-ray sources in the nearby Universe together with the {\it Fermi}/LAT\footnote{\url{https://www.slac.stanford.edu/exp/glast/groups/canda/lat_Performance.htm}} and CTA-South\footnote{\url{https://portal.cta-observatory.org/Pages/CTA-Performance.aspx}} sensitivities for 10-yr integration and a 50~hr observation, respectively. We select representative sources from \citet{cse12} which include microquasar S26 \citep{pak10}, ULX IC~342-X-1 \citep{all08,cse12}, and high-mass X-ray binary (HMXB) IC~10-X-1 \citep{loz07}. The parameters are summarized in \autoref{tab:neb}. For the calculation of the $pp$ interaction, we follow the prescription provided by \citet{kam06} setting the fraction of the energy received in the pion as 0.17 \citep{kel06}. For the gas density, we consider two cases with $n_{\rm gas}=1~{\rm cm^{-3}}$ and $3~{\rm cm^{-3}}$. Here, ULXs are known to be associated with star forming regions \citep[e.g.][]{swa09,gri11,pou13} and the gas density in the H~II region can be as high as $\sim10~{\rm cm^{-3}}$ for the size of $\sim$100~pc \citep{hun09}.

The nebula associated with the nearby microquasar S26 shows a flat spectrum in $E^2dN/dE$ extending to $\sim100~{\rm TeV}$ and is detectable even for $n_{\rm gas}=1~{\rm cm^{-3}}$. Thus, S26 would be a good target for future CTA observation. On the contrary, the nebulae associated with IC~342~X-1 and IC~10~X-1 would have soft gamma-ray spectra at $\gtrsim10~{\rm GeV}$ due to ion-neutral collisions, as the expansion velocity is slower than 120~${\rm km~s^{-1}}$. Such slowly expanding nebulae would be difficult to be observed by CTA.

\begin{figure}
 \begin{center}
  \includegraphics[width=8.5cm]{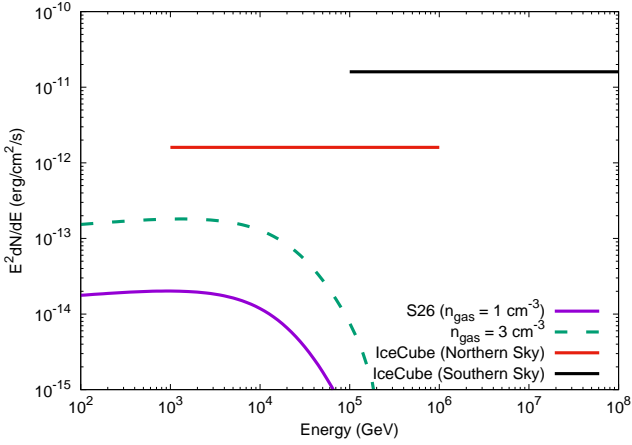} 
 \end{center}
\caption{Expected neutrino spectrum per flavor from the nebula associated with microquasar S26. The thick and dashed curve represents the case with $n_{\rm gas}=1~{\rm cm^{-3}}$ and $3~{\rm cm^{-3}}$, respectively. The sensitivities of IceCube for the northern sky and the southern sky  \citep{aar14_sens} are also shown, respectively.}\label{fig:neu}
\end{figure}

\autoref{fig:neu} shows the expected neutrino signal per flavour from the nebula surrounding S26 together with the IceCube sensitivity \citep{aar14_sens}. If the gas density is as high as $n_{\rm gas}=3~{\rm cm^{-3}}$, an order of magnitude sensitive neutrino detectors will be able to see the signal.

\citet{bor15} reported the marginal detection of gamma rays from the direction of  the Galactic microquasar SS~433 and the related supernova remnant W50 using {\it Fermi}. These gamma-ray signals may arise from the particle acceleration in the associated nebula. However, the shell expansion velocity is known to be $16-40~{\rm km~s^{-1}}$ \citep{loc07} which would be inefficient to accelerate particles in the expanding shell. 

Electrons are also expected to be accelerated together with protons. To evaluate the leptonic contribution, we need an electron-to-proton ratio $K_{ep}$, where $K_{ep}$ is defined as the ratio of electron spectrum $dN_e/dp$ and proton spectrum $dN_p/dp$ at $p = 1~{\rm GeV~c^{-1}}$. In this paper, we take $K_{ep}=0.01$ as the fiducial value, since this value is favored for Galactic SNRs in the hadronuclear scenario \citep{ack13_pion}. The expected radio synchrotron flux by electrons is given by $f_{e,\rm syn}\approx c\sigma_TU_B\gamma_{e,s}^3N_e(\gamma_{e,s})/{6\pi d^2}$ \citep{der09}. Thus, the radio flux by primary electrons  is approximately given as 
\begin{eqnarray}
\nonumber
E_\gamma^2 \frac{dN_\gamma}{dE_\gamma}&\sim&1.2\times10^{-16}~{\rm erg~cm^{-2}~s^{-1}}~\left(\frac{\mathcal{C}}{12}\right)^{-1}\left(\frac{d}{2~{\rm Mpc}}\right)^{-2}\\
&\times&K_{ep,-2}\xi_{{\rm cr},-1}R_{{\rm b}, 20.8}^3v_{s, 7.1}^{2}n_{{\rm gas},0}^{7/4}\nu_{s, 9.7}^{1/2},
\end{eqnarray}
where $\nu_s$ is a frequency of the synchrotron component. By adopting the parameters for S26 and assuming $K_{ep}=0.01$, we get $\sim2.6$~mJy and $\sim2.0$~mJy at 5.5 and 9.0~GHz, respectively, with $n_{\rm gas}=3~{\rm cm}^{-3}$ which are consistent with the measured radio flux of S26 of 2.1~mJy at 5.5~GHz and 1.6~mJy at 9.0~GHz \citep{sor10}.  Thus, determination of gamma-ray spectrum will be also useful to understand the origin of the radio emission and to constrain the environment of the nebula. If the downstream magnetic field is amplified by Alfv\'en-wave turbulence, the required $K_{ep}$ becomes $7\times10^{-5}$ to be consistent with the radio observations.

Here, the inverse Compton (IC) flux by electrons scattering the cosmic microwave background (CMB) photons is given as $f_{\rm IC}\approx c\sigma_T\gamma_{e,s}^3N_e(\gamma_{e,s})u_{\rm CMB}/6\pi d^2$, where a monochromatic radiation field assumed for the CMB and we assume the Thomson regime \citep{der09}. The IC flux is approximately
\begin{eqnarray}
\nonumber
E_\gamma^2 \frac{dN_\gamma}{dE_\gamma}&\sim&1.4\times10^{-13}~{\rm erg~cm^{-2}~s^{-1}}~\left(\frac{\mathcal{C}}{12}\right)^{-1}\left(\frac{d}{2~{\rm Mpc}}\right)^{-2}\\
&\times&K_{ep,-2}\xi_{{\rm cr},-1}R_{{\rm b}, 20.8}^3v_{s, 7.1}^{2}n_{{\rm gas},0}^{7/4}E_{\gamma, 12}^{1/2},
\end{eqnarray}
Therefore, the IC contribution from primary electrons would become important at $E_\gamma\sim1$~TeV if $K_{ep}\gtrsim0.01$. Since secondary electrons carry only $\sim0.8$\% of the primary proton energy, their contribution would be less important. If the local optical or infrared (IR) photon energy density is comparable to the CMB energy density, the total IC contribution would be enhanced. The local nebular luminosity of S26 is $\sim5\times10^{39}~{\rm erg~s^{-1}}$ \citep{dop12} corresponding to the photon energy density of $\sim0.020~{\rm eV~cm^{-3}}$. Thus, the IC flux for S26 would not be significantly enhanced even if we take into account local photon fields other than CMB photons.

\section{Discussions}
\subsection{Spectral Index of Accelerated Cosmic Rays}
In this paper, we assume the spectral index of 2 for accelerated particles. This assumption would be violated by various possible mechanisms. As the source S26 expands at the velocity of $250~{\rm km~s^{-1}}$ corresponding to the sonic and Alfv\'enic mach number of 17 and 35, respectively, we can expect the compression ratio $r\sim4$ for the sources S26 in NGC~7793. However, for sources having lower expansion velocity, the compression ratio would become smaller resulting in softer spectra \citep[e.g.][]{cap14}. In addition to this effect, the ion-neutral collision effect \citep{shu79,hol89,lee15} and the effect of finite velocity of the counter streaming magnetic turbulence in front of the shock \citep[e.g.][]{lee12,byk14} would steepen the cosmic-ray spectral index. On the contrary, in the radiative region, we expect a larger compression ratio than $r=4$. The size of the radiative region is $l_{\rm rad}\approx v_s t_{\rm rad}/4\sim4.3\times10^{-2}n_{\rm gas,0.5}^{-1}v_{s,7.1}^{22/5}~{\rm pc}$, while the size of the acceleration region is $l_{\rm acc}\approx v_s t_{\rm acc}\sim4.8\times10^{-2}Z^{-1}n_{{\rm gas}, 0.5}^{-1/2}v_{s,7.1}^{-1}E_{{\rm cr}, 10.5}~{\rm pc}$. Thus, the spectral index would become harder than 2 above $\sim30$~GeV and more photons would be detectable at the TeV band. 

\subsection{Cosmic Gamma-ray and Neutrino Background Radiation}
The cosmic GeV gamma-ray background radiation is known to be composed of blazars \citep[e.g.][]{ino09,aje15}, radio galaxies \citep{ino11_rdg}, and star forming galaxies \citep{ack12_stb}. However,  there is still an unresolved component at the flux level of $7.2\times10^{-6}\ \rm{cm^{-2}s^{-1}sr^{-1}}$ above 0.1~GeV \citep{ack15_cgb}. If nebulae associated with microquasars and ULXs emit gamma rays, they would also contribute to the unresolved cosmic gamma-ray background radiation. 

The local X-ray luminosity functions of ULXs are well studied in literature \citep[e.g.][]{gri03,wal11,swa11,min12}. The local number density of ULXs is $n_{\rm ULX}\sim1.5\times10^{-2}~{\rm Mpc}^{-3}$ \citep{swa11}. Then, the cosmic gamma-ray background flux from ULX/microquasar nebulae can be estimated as
\begin{eqnarray}
E_\gamma^2 \frac{dI_\gamma}{dE_\gamma} &\approx& \frac{1}{3}\frac{ct_H\xi_z}{4\pi}\frac{\epsilon_{\rm M}\xi_{\rm cr}P_{\rm kin}n_{\rm ULX}{\rm min}[1, f_{pp}]}{\mathcal{C}}\\ \nonumber
&\sim&5.2\times10^{-6}~{\rm MeV~cm^{-2}~s^{-1}~sr^{-1}}~\left(\frac{\xi_z}{3}\right)^{-1}\left(\frac{\mathcal{C}}{12}\right)^{-1}\\
&\times&\xi_{{\rm cr},-1}R_{{\rm b}, 20.8}^3v_{s, 7.1}^{2}n_{{\rm gas},0.5}^2n_{{\rm ULX},-1.8},
\end{eqnarray}
where $t_H\sim13.8~{\rm Gyr}$ is the Hubble timescale assuming $H_0=67.8~{\rm km~s^{-1}~Mpc^{-1}}$ and $\Omega_M=0.308$ \citep{ade16}. $\xi_z$ is a factor accounting cosmological evolution of the source. We assume $v_s\ge120~{\rm km~s^{-1}}$ for all sources. Therefore, ULX/microquasar nebulae would contribute $5.2\times10^{-7}\ \rm{cm^{-2}s^{-1}sr^{-1}}$ above 0.1~GeV ($\sim7\%$ of the unresolved background flux). 

The cosmic neutrino background flux would become 
\begin{eqnarray}
\nonumber
E_\nu^2 \frac{dI_\nu}{dE_\nu} &\sim&2.6\times10^{-9}~{\rm GeV~cm^{-2}~s^{-1}~sr^{-1}}~\left(\frac{\xi_z}{3}\right)^{-1}\left(\frac{\mathcal{C}}{12}\right)^{-1}\\
&\times&\xi_{{\rm cr},-1}R_{{\rm b}, 20.8}^3v_{s, 7.1}^{2}n_{{\rm gas},0.5}^2n_{{\rm ULX},-1.8},
\end{eqnarray}
where we note that the neutrino spectra from ULX nebulae would have a cutoff at several tens TeV (see \autoref{eq:Emax}). Due to this cutoff, the nebulae would not contribute to the IceCube detected neutrino fluxes \citep{aar14}. However, if Alfv\'en-wave turbulence at the shock is fully generated, sub-PeV neutrinos can be affordable.

\subsection{Gamma-ray Emission and Attenuation in Host Galaxies}
The total IR luminosity of NGC~7793 which hosts the microquasar S26 is $2.07\times10^9L_\odot$ \citep{gal13}. Adopting the scaling relation between gamma-ray and IR luminosities \citep{ack12_stb}, we have $L_\gamma\sim3.8\times10^{38}~{\rm erg~s^{-1}}$ for galactic diffuse emission leading $E_\gamma dN_\gamma/dE_\gamma\sim2.2\times10^{-13}~{\rm erg~cm^{-2}~s^{-1}}$ which is comparable to expected flux from S26 assuming $n_{\rm gas}=1~{\rm cm^{-3}}$. Here, the angular size of NGC~7793 is $9'3\times6'3$ and S26 locates at the outskirt region of the galaxy ($\sim3$~arcmin away from the nucleus). Given the CTA angular resolution of $\lesssim3$~arcmin at $\gtrsim1$~TeV, gamma-ray signals from S26 can be distinguished from gamma rays due to galactic cosmic rays which would be bright in the star forming nucleus region.

It is well-known that gamma rays propagating the intergalactic space are attenuated by the cosmic optical/IR background radiation field via the pair production process \citep{gou66,jel66,ste92}. Since we are considering the local universe $\lesssim10$~Mpc, the gamma-ray attenuation by the cosmic optical/IR background is negligible \citep[e.g.][]{ino13}. However, if ULXs and microquasars are formed in the star forming regions, internal gamma-ray attenuation by the interstellar radiation fields will suppress gamma-ray flux above 10~TeV \citep[see e.g.,][]{ino11}.

\subsection{ULX pulsars}
Recently, some ULXs are discovered to be hosted by accreting pulsars \citep{bac14,fue16,isr16}. Physical properties of ULX pulsars are still open questions. Magnetic fields of M82~X-2 are ranging $10^9~{\rm G}\lesssim B\lesssim 10^{14}~{\rm G}$ in literatures \citep{ekc15,klu15,mus15,tsy16,kin16,kar16,che16}.

If we assume that X-ray emissions from ULX pulsars are generated at accreting column \citep[e.g.][]{kaw16}, the accretion shock may accelerate particles near the polar cap region of the pulsars. However, $\gtrsim10$~GeV gamma rays from the polar cap region emission will be attenuated due to one-photon pair creation. The threshold magnetic field strength is $B\gtrsim 0.1B_Q$ \citep{har06,dau82}, where $B_Q = m_e^2 c^3/e = 4.4\times10^{13}$~G. Therefore, strong gamma-ray emission would not be expected for high magnetic field ULX pulsars.

\section{Conclusions}
In the coming decade, CTA will deepen our understandings of high energy astrophysical phenomena. In this paper, we investigate high energy signals from nebulae associated with microquasars and ULXs in the nearby Universe whose available power is $\gtrsim10^{39}~{\rm erg~s^{-1}}$. 

Although gamma rays from host galaxy itself exist, CTA's sensitivity and angular resolution allow us to detect and resolve powerful nebulae associated with microquasars and ULXs. For example, the nebula associated with the microquasar S26 in NGC~7793 \citep{pak10} will be a good target whose expected gamma-ray flux is $\sim1\times10^{-13}$--$1\times10^{-12}~{\rm erg~cm^{-2}~s^{-1}}$ extending to several tens TeV with a photon index of $\sim2$. At $\gtrsim1$~TeV, the IC contribution would become important if $K_{ep}\gtrsim0.01$. We note that the spectral index of cosmic rays would be harder than 2 above $\sim30$~GeV since it is the radiative shock. We also found that the expected synchrotron radiation flux from primary electrons is consistent with the measured radio flux. If all microquasars and ULXs are associated with powerful expanding nebulae, they would make $\sim7$\% of the unresolved cosmic gamma-ray background flux. However, they would not significantly contribute to the TeV-PeV neutrino background flux reported by IceCube due to cutoffs at several tens TeV in neutrino spectra. If the expansion velocity is $v_s\lesssim120~{\rm km~s^{-1}}$, ion-neutral collisions will suppress efficient particle acceleration and result in fainter TeV gamma-ray flux above $\sim10$~GeV.


\subsubsection*{Acknowledgements:} 
The authors thank Chris Done for useful comments and discussions. YI is supported by the JAXA international top young fellowship and JSPS KAKENHI Grant Number 420 JP16K13813.

\end{document}